\documentclass[submission,copyright,creativecommons]{eptcs}
\providecommand{\event}{FROM 2026} 
\usepackage{multicol}
\usepackage{amsmath}
\usepackage{iftex}
\usepackage{tikz}
\usepackage{xcolor}
\usepackage{listings}
\usepackage{amssymb}
\usepackage{xurl}
\usetikzlibrary{automata, positioning, arrows.meta, calc}
\ifpdf
  \usepackage[strings]{underscore}         
  \usepackage[T1]{fontenc}        
\else
  \usepackage{breakurl}          
\fi

\definecolor{vscodeKeyword}{HTML}{0000FF}   
\definecolor{vscodeType}{HTML}{267F99}      
\definecolor{vscodeFunc}{HTML}{795E26}      
\definecolor{vscodeString}{HTML}{A31515}    
\definecolor{vscodeComment}{HTML}{008000}   
\definecolor{vscodeNumber}{HTML}{098658}   
\definecolor{vscodeBg}{HTML}{FFFFFF}        
\definecolor{vscodeLineNo}{HTML}{237893}    
\providecommand{\keywords}[1]
{
  {\small	
  \textbf{\textit{Keywords---}} #1}
}

\lstdefinelanguage{Dafny}{
  morekeywords=[1]{
    datatype, class, module, type, function, method, predicate,
    requires, ensures, reads, modifies, decreases, invariant,
    var, const, return, if, else, while, for, match, case,
    assert, assume, new, this, true, false, null,
    forall, exists, in, out, ghost, static, import, export,
    lemma, returns, yield, break, continue, then, by
  },
  morekeywords=[2]{
    int, bool, real, char, string, nat, object, array, seq,
    set, map, multiset, Option, Symbol, InputSymbols,
    AdditionalTapeSymbols, Conclusion, State
  },
  sensitive=true,
  morecomment=[l]{//},
  morecomment=[s]{/*}{*/},
  morestring=[b]",
}

\title{The Formalization of two Computational Models in Dafny\\
\large Short Paper}
\author{\c{S}tefan Ciob\^{a}c\u{a} \qquad\qquad Diana-Elena Gratie \qquad\qquad Drago\c{s}-Irinel Rotariu {
  
  \footnote{Authors in alphabetical order.}
}
\institute{Alexandru Ioan Cuza University of Ia\c{s}i, Romania}
\email{stefan.ciobaca@uaic.ro \qquad\qquad diana.gratie@uaic.ro \qquad rotariudragos36@gmail.com}}

\def\titlerunning{The Formalization of two Computational Models in Dafny}
\def\authorrunning{\c{S}. Ciob\^{a}c\u{a}, D.E. Gratie, D.I. Rotariu}

\begin{document}
\maketitle

\begin{abstract}
    We describe the formalization in Dafny of two computational models, Turing Machines and the Lambda Calculus. We present several application of the formalizations: machine proofs of termination for Turing machines, Dafny proofs for Church encodings, and a mechanized proof of the Church-Rosser theorem.
\end{abstract}


\section{Introduction} 

The ability to solve a problem using an effective procedure \cite{computability} or computability can lead to a concrete, mathematical way to prove the complexity of problems, or it can explain the properties of different programming languages and their differences when it comes to evaluation (an example is the lazy versus eager evaluation debate). Still, using mathematical theory alone may not be enough for some people. As such, a proper formalization of the mathematical definitions surrounding computational models raises the level of understanding and trust surrounding them and their properties.

This paper explores the formalizations of two well-established computational models, Turing Machines and Lambda Calculus, in Dafny. 
Dafny~\cite{leino2010dafny} is a multi-paradigm, verification-aware programming language offering built-in support for recording specifications and a static program verifier. The main way to formally prove that a user-defined function behaves properly is to use preconditions (introduced by $\texttt{requires}$), postconditions (introduced by \texttt{ensures}), make assertions, and make lemmas to aid the prover in verifying your code. 
Dafny thus is a very useful tool for writing provably correct code (with respect to the given specifications), making rigorous verification an integral part of development and reducing bugs. The language includes bounded and unbounded quantifiers, the ability to use and prove lemmas, user-defined mathematical functions and ghost variables  and predicates. For more details on the Dafny programming language, we refer to~\cite{dafny}.

Dafny is the perfect middle ground between classic programming languages like Java, Python, Rust, and Haskell, and interactive theorem provers like Rocq (formerly known as Coq)\cite{rocq} and Lean \cite{lean}. Dafny's way of proving that the code works in a mathematical way by using preconditions, post-conditions and lemmas while maintaining the main code in a  practical state, like any other programming language, makes Dafny the perfect choice for the formalization of the computational models. 

In Section~\ref{sec:contrib} we present our contributions and related work. Section~\ref{sec:TM} presents the formalization of Turing Machines, and is mainly based on the \emph{Mathematical Foundations of Computing} course offered by Stanford University \cite{stanford:cs103} and the \emph{Computability, Decidability and Complexity} course offered by the Faculty of Computer Science in Ia\c{s}i \cite{tiplea:cdc}. It also discusses some related work. Section~\ref{sec:LC} presents the Lambda Calculus, its formalization, as well as the Church encoding, which are a recreation of the Haskell implementation of the Lambda Calculus and Beta Reduction from the Functional Programming course offered by the Faculty of Computer Science in Ia\c{s}i~\cite{course:fp}. Section~\ref{sec:futureWorks} gives possible future work based on our projects and we conclude in Section~\ref{sec:conclusions}.
    
\section{Contributions}\label{sec:contrib} 
There are many articles that tackle the problem of formalizing the two selected computational models, with differences in terms of the selected model (e.g., Deterministic vs. Nondeterministic TM) and theorem prover. The more mathematical nature of ordinary theorem provers may lead to a harder time for a programmer that looks for a more practical explanation of computational models to be able to understand their hidden mechanics. 

To be more specific, while there are articles that explore the topic of Turing Machines formalizations~\cite{formalizingTM, lederer2026verify} using various provers~\cite{asperti2011matita} including Dafny, they use Deterministic Turing Machines instead of Nondeterministic Machines. Our formalization takes into account both variants of a TM and makes the differences and similarities between these two clear.

There also exist several formalizations~\cite{nipkow1996more, Mechanical_proof, huet1994residual} for the Lambda Calculus, typically with applications in proving meta-theorems using assistants such as Isabelle/HOL~\cite{isabelle_website}, The Boyer-Moore Theorem Prover~\cite{boyer_moore_nqthm}, or Rocq~\cite{rocq}.

Most use the de Bruijn indices to represent variables rather than the text book named variables. The de Bruijn index is very useful and simplifies many problems but it may not be as intuitive for novices as the standard named variables; some mathematical proofs also use named variables when using Lambda Calculus so we opted to use that too.

Thus, this project raises the understanding of Lambda Calculus and Turing Machines by strictly formalizing the standard theoretical foundations without omitting key elements that may or may not be necessary for proving theorems but are crucial in making the foundations of the formalization. 

This is why our formalization is mainly  a methodological and theoretical contribution to the literature as it sheds light upon some concepts that are rather important in understanding those computational models and computational theory while using a formal verifier that is closer to an actual programming languages than standard theorem provers.

For some parts of the formalizations, we have used AI. We discuss the experience and we analyze the capabilities of conversational and autonomous agents when it comes to formal verifications and formally proving parts of the theorem as a case study about the use of AI in formalizations.

The source code of the project can be found on Github~\cite{rotariu_github}. This paper is based on the BSc thesis of the third author~\cite{rotariu_bsc_thesis}.

\section{Turing Machines and their formalization in Dafny}\label{sec:TM}
     We formalize a textbook~\cite{tiplea:cdc} representation of Turing Machines (TMs), as a tuple $M = (Q, \Sigma, \Gamma, \delta, q_0, \_, F)$ where: 
    \begin{itemize}
        \item $Q$ is a finite non-empty set of states
        \item $\Sigma$ is a finite non-empty set of input symbols named the input alphabet of M 
        \item $\Gamma$ is a finite non-empty set of symbols that can be found in the tape, with $\Sigma \subset \Gamma $
        \item $\delta: Q \times \Gamma \rightarrow 2^{Q \times \Gamma \times \{L, R\}}$ is the transition function of $M$
        \item $q_0 \in Q$ is a state called the initial state of $M$
        \item $\_ \in \Gamma - \Sigma $ is the symbol that represents the empty cell in the tape, called blank symbol
        \item $F \subseteq Q $ is a set of final states
    \end{itemize}

A Turing Machine can be Deterministic (DTM) or Nondeterministic (NTM), depending on whether there is only one action ($Q \times \Gamma \times \{L, R\}$) for every key ($\delta: Q \times \Gamma$) or not. 

We use an algebraic data type for \texttt{Symbols}, with it being a blank or non-blank symbol that wraps a string. We have a clear distinction between DTM and NTM based on the transition relation $\delta$. We use the name $\mathit{transitions}$ for the NTM and the name $\mathit{deterministicTransitions}$ for the DTM.

We define $\mathit{State}$ as a name for the state name, plus a conclusion (Accept or Reject) wrapped in an option:  ($Option<Conclusion>=\mathit{Some}(Conclusion:c)|None$).

We define the tape as a function that takes a number and returns a $\mathit{Symbol}$, enabling the representation of unbounded tapes. 

This technique is possible because most of the cells are empty, and thus we can use a function that retains only the non-empty cells and returns $\mathit{Blank}$.

A configuration of a Turing Machine stores the state that it is in, the position of the head, and the tape. 

The last thing that needs to be addressed is the representation of the transition relation, $\delta$. As suggested above, there is a need for: 
\begin{itemize}
    \item A key that is the combination of a state and the symbol at which the head is pointed
    \item An action that represents a step of the Turing Machine (modifies the cell symbol being pointed at, changes the state, and moves to the nearest cell to the left or right).
\end{itemize}

The need for a different data structure is caused by two reasons. The first reason is the difference between NTM and DTM in terms of mathematical definition  \cite{tiplea:cdc}, and the second reason is the overcomplication of working with a DTM that is represented through an NTM.

Even though in theory all DTMs are NTMs and it is possible to use an NTM to represent a DTM, in practice and in the code, those two are represented through different data structures, so the link is not clear for the verifier. 
Thus, we need to formally prove this link between them, which is one of the challenges in this project. To link those two different data structures, we define a predicate that checks whether a transition relation $\mathit{Transitions}$ can be translated into a function $\mathit{DeterministicTransitions}$ (for all keys, there is only one action available) and two functions that can transform an NTM into a DTM and vice versa (by transforming the corresponding $\delta$). 

This establishes the above goal. Moreover, most of the functions and predicates with NTMs have a DTM counterpart that must be linked by lemmas proving their similar functionality. This process helped us better understand the process of linking similar concepts and emphasizes the major impact that using two data structures that are similar but still different can have in formal proofs.

A step in a TM represents the transition from a configuration to another one by getting the key from the configuration and applying a specific action that corresponds to the key according to $\delta$ (and a position if it is an NTM). If there is no key in $\delta$, the TM will fail as in real life by giving a $\mathit{None}$  instead of $\mathit{Some}(\mathit{configuration})$ (like a $\mathit{conclusion}$ of a $\mathit{State}$).

The need of a position for the NTM makes the function that simulates an NTM rather demanding and complex. That is why we should use a DTM to represent a TM: to simplify and streamline the simulation of a TM, as it is possible to simulate any NTM by using a modified DTM with a certificate \cite{stanford:cs103} (the proof is out of the scope of the paper).

A major problem when formally proving algorithms and functions in general is proving the termination of a function, as it needs a clear indication of an element that is decreasing in a way until it reaches a base case where the function terminates. This is the same case in Dafny too, with the added fact that a TM can loop forever complicating things further. 
To model termination of TMs, we rely on the following technique: we use a predicate that ensures the existence of a finite path. More precisely, a path of steps (or transitions) between two configurations such that the path has an exact number of steps that is given to us. That natural number is the needed indication of getting closer to the final configuration, with it reaching 0 signaling the end of the path.

\begin{lstlisting}[basicstyle=\ttfamily\scriptsize\color{black}]
ghost predicate isThereAClosedTransitionInNSteps
(delta:Transitions, conf1:Configuration, conf2:Configuration, n:nat) 
    decreases n
{
  if n==0 then conf1==conf2
    else  
  exists conf',poz:nat::(isPozInTransitions(conf1,delta,poz)) 
  && applyTransition(conf1,delta,poz)==Some(conf') 
  && isThereAClosedTransitionInNSteps(delta,conf',conf2,n-1) 
}
\end{lstlisting}

To extract the number $n$, we need a predicate that confirms its possible existence. This method of guaranteeing the existence of an element that indicates the eventual termination of the function is used frequently throughout the whole project and will also be used for future work. 

\begin{lstlisting}[basicstyle=\ttfamily\scriptsize\color{black}]
ghost predicate isThereAClosedTransition
(delta:Transitions,conf1:Configuration, conf2:Configuration)
{
  exists n:nat:: isThereAClosedTransitionInNSteps(delta, conf1, conf2, n)
}    
\end{lstlisting}

Now, to determine if an $\mathit{input}$ halts, one needs to find a finite, well-defined path of transitions from the $\mathit{initialConfiguration}$ generated by it to a $\mathit{Configuration}$ that has a final state. The conclusion of that final state indicates whether the input is accepted or rejected. 

\begin{lstlisting}%[basicstyle=\ttfamily\tiny\color{black}]
[basicstyle=\ttfamily\scriptsize\color{black}]
ghost predicate isAcceptedInTM 
(delta : Transitions, q0:State,
inputS:InputSymbols,addTapeS:AdditionalTapeSymbols,
input:seq<string>)
  ...\\needed requirements
{
  exists conf:Configuration :: isConfAccepted(conf) && 
  isThereAClosedTransition(delta,initialConfiguration(input,q0,inputS),conf)
}
    
\end{lstlisting}

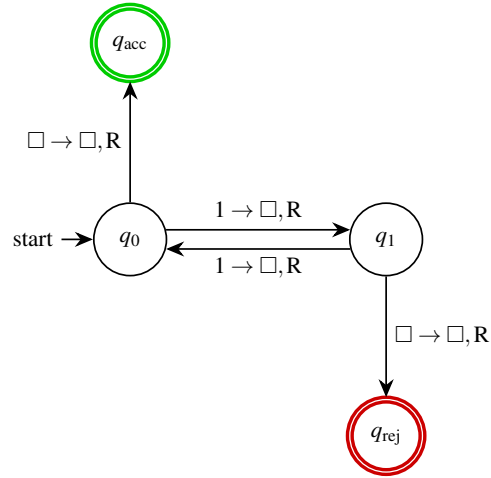
\begin{figure}[h!]
    \centering
\scalebox{0.8}{
\begin{tikzpicture}[
    > = {Stealth[length=3mm, width=2.5mm]},
    thick,
    node distance=2.5cm and 3cm, 
    every state/.style={thick, fill=white, minimum size=1.2cm},
    accept/.style={accepting, draw=green!80!black, fill=white, ultra thick},
    reject/.style={accepting, draw=red!80!black, fill=white, ultra thick}
]

\node[state, initial, initial text=start, initial where=left] (q0) {$q_0$};
\node[state, right=of q0] (q1) {$q_1$};
\node[state, accept, above=2cm of q0] (qacc) {$q_{\text{acc}}$};
\node[state, reject, below=2cm of q1] (qrej) {$q_{\text{rej}}$};

\path[->]
    (q0) edge node[left] {$\square \to \square, \text{R}$} (qacc)
    (q1) edge node[right] {$\square \to \square, \text{R}$} (qrej);

\draw[->] (q0.15) -- (q1.165) node[midway, above] {$1 \to \square, \text{R}$};
\draw[->] (q1.195) -- (q0.345) node[midway, below] {$1 \to \square, \text{R}$};

\end{tikzpicture}
}
    \caption{A simple Turing Machine}
\end{figure}

With the number of steps revealed, it is possible to determine if a TM halts in polynomial time by comparing the length of the input with the number of steps: $\exists m. \space  n=|input|^m$, with $n$ being the number of steps.

This also further helps with future proofs that a problem is Nondeterministic Polynomial ($\mathit{NP}$) by proving that the language associated to it is accepted (all inputs in it are accepted) by an NTM (that will be simulated through a DTM). 

The first application of this paper is proving that the TM in Figure 1 halts. The TM is one that receives a string of $1$s, accepts it if the input's length is even, and rejects it otherwise. 

In the main method, we proved that the process of transforming the $\mathit{Transitions}$ into \\$\mathit{DeterministicTransitions}$ is correct and then it halts for every input by linking the intermediary configurations with the initial and final steps. 

\pagebreak
\begin{lstlisting}%[basicstyle=\ttfamily\tiny\color{black}]
[basicstyle=\ttfamily\scriptsize\color{black}]
lemma CorrectInputHalts(
  delta: DeterministicTransitions, 
  inputS: InputSymbols, 
  addTapeS: AdditionalTapeSymbols, 
  q0: State, 
  q1: State, 
  qRej: State, 
  qAcc: State,
  one: Symbol, 
  input: seq<string>
)
  requires isInputCorrect(input)
  ...// is input correct, is TM defined exactly as the figure
  ensures haltsInDTM(delta, q0, inputS, addTapeS, input)
{
  var init_config := initialConfiguration(input, q0, inputS);
  var steps := |input|;

  var mid_config := TraceExecution(delta, inputS, addTapeS, init_config, steps);
  
  match applyTransitionDTM(mid_config, delta)
   case Some(halt_config) => {
    LinkTransitions(delta, inputS, addTapeS, init_config, mid_config, halt_config, steps);

    assert isThereAClosedTransitionInNStepsForDTM(delta, inputS, addTapeS, init_config, halt_config, steps + 1);
    assert isThereAClosedTransitionDTM(delta,inputS,addTapeS,init_config,halt_config);
    assert haltsInDTM(delta, q0, inputS, addTapeS, input);
    }}
\end{lstlisting}

\section{Lambda Calculus and its formalization in Dafny}\label{sec:LC}

A lambda-term (or simply referred as term) can be: 
\begin{itemize}
    \item a variable (element from the $\mathit{Id}$ set of identifiers),
    \item a lambda-abstraction ($\lambda x. t$ with $x$ being a variable and $t$ being another term),
    \item or a lambda-application ($t_1 \space t_2$, with $t_1$ and $t_2$ being terms).
\end{itemize}

We formalize lambda-terms as algebraic types, where variables are identifiers represented by natural numbers. Because of this representation, most functions also take a set of the free variables and a set of bound variables as argument.

Two functions that need to be pointed out are substitution and capture-avoiding substitution: The usual substitution, $t[x/t']$, simply replaces every free instance of the variable $x$ in the term $t$ by the term $t'$. This is susceptible to name capturing, as it has the possibility of making formerly free variables bound by changing the variable to one that was already bound by an abstraction, for example: \\ $(\lambda x.y)[y/x]=\lambda x.x$.

The capture-avoiding substitution (or CA substitution), $t[[x/t']]$ , solves this issue by alpha-renaming in the case of $t=\lambda y. t_1$, where $y$ is found in $free(t')$ :  $(\lambda y.t)[[x/t']]=\lambda y'.((t[y/y'])[x/t']) \space, y\ne x$, with $y'$ being a fresh variable that is not found in neither $t$ nor $t'$.

Formalizing the capture-avoiding substitution presents two difficulties:
\begin{itemize}
    \item the termination argument is not clear;
    \item the generation of fresh variable names.
\end{itemize}

The first problem in $t[[x/t']]$ rises because of the substitution of the violating variable that we do in the case of $t=\lambda y.t_1$
where $y \in \mathit{free}(t_1)$. In the capturing substitution it just went deeper in the term, similar to the traversal in a tree, but in the CA substitution, instead of going into $t_1[x/t']$, it will go into $t_1[y/y'] [[x/t']]$ which complicates things. 
Fortunately, the height of a term is enough of the function decreasing as it can be proved that $t_1[y/y']$ and $t_1$ have the same height. Still, the verifier is not powerful enough to catch on this detail, so we needed to make a separate lemma in order to prove this statement.

For the more challenging problem: to guarantee the freshness of $y'$, there needs to be a way to retain the set of all variables in a data structure and from it to generate the new variable (that will also be added to it as it will be part of the new term). 
In the early builds of the project, we used a string to represent a variable. The main problem with this approach is that we need to generate, in a deterministic manner, a new variable in the certain edge case, and there were no viable strategies to get that never-seen variable.

Thus, we changed the representation of variables. Here is where the detail of variables being natural numbers comes into play by making these observations: 1) for any natural number $n$, there is a natural number $m=n+1$ that is greater than $n$; 2) if a number is greater than the highest number in a set, then that number is greater than any number in that set, and so it is not in that set. With both observations in mind, it is possible to generate a fresh variable by finding the highest variable number in $\mathit{ids}$ and then increasing it by one. 

The difficulties do not end here as $\mathit{ids}$ cannot be defined as a set without rendering the whole function viable only for verification, as there is no deterministic way to get the highest number from a set. 
The reason behind this restriction is that Dafny does not allow the extraction of a specific element from a set (even if it only has a single element). To make it valid for use in practice, some tinkering was needed, such as making the function that gives all the variables in a set, $\mathit{var}$, return a list instead of a set.

To comply with the observations above, the list will have the behavior of an annotated set: all elements are unique along with functions that imitate the classical operations of sets ($\subseteq$, $\cup$, $-$). 
With this and other helper lemmas that ensure that the $\mathit{ids}$ used in the recursion are correct, the capture-avoiding substitution can be fully built. This whole process of carefully choosing which data structure we need to use and how to simulate mathematical structures will prove to be useful experience for the future.

Alpha-equivalence is a binary relation between two lambda-terms that holds true if the two terms are structurally the same, with the only difference being the variable names of some abstractions. This allows one term to be transformed into the other term by simply using alpha-conversions.  That was the initial idea: to make multiple alpha-conversions by using capture-avoiding substitution, but this made the predicate extremely expensive and not ideal for a function that will be heavily used.

Thus, another approach was needed. Alpha-equivalent terms are structurally the same and have the same free variables. The only thing that may differ is the variable naming of the bound variables. 
One case that can be safely worked is: if one term deviates from the other (one is an abstraction and the other is not, for example), that guarantees that the terms are not alpha-equivalent. The analysis of the variables becomes way more tricky with the recursion, but with two lists to dynamically save the bound variables in the order they appear in the abstraction as the predicate traverses through the two terms, one can analyze the bound and free variables of each term. 

For the variables, we need to see if the variables are bound by a higher abstraction that is retained in the lists mentioned above: 
\begin{itemize}
    \item if one variable $x$ is bound by an abstraction with the index $n$ in the list, then the other variable $x'$ must also be bound by the abstraction with the same index in their list of bounded variables for the terms to be alpha-equivalent.
    \item if that variable $x$ is not bound by any abstraction, then it is a free variable and the other variable $x'$  must be free and equal to it. 
\end{itemize}

The abstraction that corresponds to the variable is the leftmost appearance of it in the bound list (as the innermost abstraction has priority when it comes to bounding). This with the structural equivalence ensures that if the two terms are alpha-equivalent, the only difference is the naming of the variables in the abstraction.  

There are many approaches to the same function, but some may be more costly than others. In formalization, there needs to be a degree of optimization for functions that are used frequently so that the verifier can confirm the correctness of a program in a rather decent time.

The only rule that defines a step in Lambda Calculus is the beta-reduction:  $(\lambda x.t) \space t' \rightarrow _\beta t[[x/t']]$. A beta-reduction is applied only if there are any $\beta$-redexes in the term, and a full beta-reduction stops where there are no $\beta$-redexes to reduce. As such, when we make a reduction, we also need to point out which $\beta$-redex we want to reduce.
For that, a theoretical indexing of all $\beta$-redexes is applied from the left outermost to the right innermost, with the condition that the index used must be smaller than the number of $\beta$-redexes in the term. This facilitates the finding of the $\beta$-redex based on the index. So, in the $betaReductionStep$ function, the search for the $\beta$-redexes is made by using the index and determining in which subterm it is.

The first application with Lambda Calculus is a mechanical proof of Church encodings, where we have mechanically verified that the logical operators have their expected behavior (example: ((AND TRUE) TRUE) reduces to true).

The first step in this proof is to make a predicate that verifies if the term is alpha-equivalent to a certain boolean value or logical operator ($\mathit{isAND}$, $\mathit{isTrue}$, etc.), a function that returns a term alpha-equivalent to that boolean value or logical operator ($\mathit{trueVar}(x,y)$, $\mathit{andVar}(x,y)$), and a function that returns the term directly with default variable.

For a full beta-reduction, a variant of it that always takes the left outermost $\beta$-redex and it stops when there is no $\beta$-redex left was used. This specific full beta-reduction strategy is called normal order and to prove termination, the same trick used in the proof that a Turing Machine halts was utilized. 

The proof for the expected behaviour of the logical operators is pretty simple: one only needs to link every normal order reduction step. As an example, this is the proof that ((AND TRUE) TRUE) reduces to TRUE:

\begin{lstlisting}[basicstyle=\ttfamily\scriptsize\color{black}]
lemma andTrueTrue(and:LambdaTerm, tru:LambdaTerm) 
    requires and==andVal() && tru==trueVal()
    ensures var andtruetrue:=Application(Application(and, tru), tru);
    normalOrderHalts(andtruetrue) && isTrue(normalOrder(andtruetrue))
{
    var s0:= Application(Application(and, tru), tru);
    var help1:= Lambda(1, Application(Application(tru, Var(1)), tru));
    var s1:= Application(help1, tru);
    var s2:= Application(Application(tru, tru), tru);
    var s3:= Application(Lambda(1, tru), tru);

    assert normalOrderStep(s0)== Some(s1);
    assert normalOrderStep(s1)== Some(s2);
    assert normalOrderStep(s2)== Some(s3);
    assert normalOrderStep(s3)== Some(tru);
    assert trueVal()==tru;
    assert normalOrderStep(tru) == None;


    assert normalOrder'(s0,4) == normalOrder'(s1,3);
    assert normalOrder'(s1,3) == normalOrder'(s2,2);
    assert normalOrder'(s2,2) == normalOrder'(s3,1);
    assert normalOrder'(s3,1) == normalOrder'(tru,0);
    assert normalOrder'(tru,0) == tru;
    assert normalOrder'(s0,4) == tru;
    
    forall m:nat | normalOrderEndsInNSteps(s0, m)
        ensures normalOrder'(s0, m) == tru
    {
        normalOrderStepsUnique(s0, 4, m); 
    }
    assert normalOrder(s0) == tru;

}
\end{lstlisting}

Another application for the Lambda Calculus is a proof of the Church-Rosser Theorem. The Church-Rosser Theorem proves the confluence of the beta-reduction. It states that: 
    \begin{center}
        let $t$, $t_1$, and $t_2$ be lambda-terms such that $t \rightarrow_\beta^* t_1$ and $t \rightarrow_\beta^* t_2$. Then there exists a lambda term $t_3$ such that $t_1 \rightarrow_\beta^* t_3'$ ,$t_2 \rightarrow_\beta^* t_3''$, $t_3=_\alpha t_3'$, and $t_3=_\alpha t_3''$
    \end{center}
    In other words the reflexive, transitive closure of beta reduction, $\rightarrow_\beta^*$, has the diamond property (which is equivalent with beta-reduction being confluent) \cite{uwaterloo:cr_thm}.
    
The proof in Dafny closely follows the lecture notes at the University of Waterloo~\cite{uwaterloo:cr_thm}: there are 7 main lemmas and the creation of a new type of reduction, $\twoheadrightarrow$, that is a parallel-reduction in which certain $\beta$-redexes can be reduced together at once. 
We then prove that $\twoheadrightarrow^*$ has the diamond property and then bridge it to the $betaReductionClosure$ to be able to prove that $\rightarrow_\beta^*$ has the diamond property.

The most interesting thing that this attempt exposed is the fundamental difference between the equal used in mathematics and alpha-equivalency and the liberal use of alpha-conversions. The na\"{\i}ve approach of using the traditional equal was unsuccessful and clearly not what the authors intended as they used an alpha-conversion on a term to escape some undesired cases but they still used the $=$ even though they are not equal terms, they are alpha-equivalent terms.
The use of alpha-equivalence as the fundamental relation between two terms proved to be very difficult as Dafny does not know much about alpha-equivalence, thus it needs many helper lemmas to aid in the proof. 

Still, this is not the only problem. The main and largest problem encountered is the alpha-conversion: $\mathit{caSubstitution}$ uses a rather rigid $\mathit{ids}$ to determine what variable names are not allowed. To alleviate this issue, one needs to prove that the use of a larger ban list for the recursive version of $\mathit{caSubstitution}$ results in an alpha-equivalent term to the original one. Also, for more flexibility, the use of alpha-equivalence should be as close and as flexible as equal. 

This and many other lemmas are needed to prove the main lemmas and even after proving them, you need to bridge $\twoheadrightarrow$ to $\twoheadrightarrow^*$ and then to bridge $\twoheadrightarrow^*$ to $\rightarrow_\beta^*$ to prove that $\rightarrow_\beta^*$ has the diamond property. This part of the proof was only hinted at in the mathematical proof \cite{uwaterloo:cr_thm}.

This application reinforces the fact that typical mathematical proofs can omit many formal details that can be cumbersome to handle formally.

\section{Future work}\label{sec:futureWorks}

There are several exciting possibilities to extend the two formalizations. For Turing Machines, it is possible to formalize the notion of NP-completeness, and, as an application, allow to formally prove NP-completeness of Dafny predicates.

For the Lambda Calculus formalization, it is possible to explore Church encodings in depth. Finally, it would also be possible to prove that Lambda Calculus are equivalent to Turing Machines. One can also study the different ways that a program is evaluated by analyzing different beta-reduction strategies and what advantages and disadvantages each one of them brings.

\section{Conclusion}\label{sec:conclusions}

The formalization of the two computational models allows to gain a deeper understanding. The formalization allows for a number of interesting applications that prove its usefulness and there are several avenues for future work, described above.

Dafny is a great tool not only to verify programs but also to formally prove not only data structures and mechanics but also complicated theorems, making it an easy and viable theorem prover for programmers that want to use a more intuitive and clear formal verifier while not being forced to learn the more complicated syntax of classical theorem provers.

Artificial Intelligence agents, more precisely Google Gemini and Claude AI, were used in this project in conversational mode. We have used Claude AI as an autonomous agent for formalizing the Lambda Calculus. 
From that experience we conclude that agentic AIs are very effective in doing concise and clear tasks like doing some helper lemmas but they often do not use the whole context of the code, that leading to many unnecessary functions or solutions that are not optimal but do the task. They also have a very hard time understanding abstract topics, thus needing to make more details and point out parts of code that already solve a problem that they observe.

\nocite{*}
\bibliographystyle{eptcs}
\bibliography{generic}
\end{document}